\documentclass[useAMS]{nature}
\usepackage{hyperref} 
\usepackage[sort,numbers]{natbib}
\usepackage{xcolor}
\usepackage[normalem]{ulem}
\usepackage{amsmath}
\usepackage{mathtools}  
\usepackage{graphicx}
\makeatletter
\let\saved@includegraphics\includegraphics
\AtBeginDocument{\let\includegraphics\saved@includegraphics}

\makeatother

\hypersetup{
    colorlinks=true,
    linkcolor=blue,
    filecolor=magenta,      
    urlcolor=cyan,
    citecolor=blue
}

\newcommand{\araa}{Annu. Rev. Astron. Astrophys.}   
\newcommand{\aj}{Astron. J.}   
\newcommand{\apj}{Astrophys. J.}   
\newcommand{\apjl}{Astrophys. J. Lett.}   
\newcommand{\apjs}{Astrophys. J. Suppl. Ser.}   
\newcommand{\aap}{Astron. Astrophys.}   
\newcommand{\mnras}{Mon. Not. R. Astron. Soc.}   
\newcommand{\nat}{Nature} 
\newcommand{\pasp}{Publ. Astron. Soc. Pac.}   

\title{Evidence of radial-migration driven Galactic disc expansion with the U-shape stellar age profile}

\author{Jianhui Lian$^{1,2\ast}$, Qinhao Shao$^{1}$, Yuze Zhao$^{1}$\\
\small $^{1}${South-Western Institute for Astronomy Research, Yunnan University, Kunming, Yunnan 650091, People’s Republic of China}\\
\small $^{2}${Yunnan Key Laboratory of Survey Science, Yunnan University, Kunming, Yunnan 650500, People’s Republic of China}\\
\small $^\ast${Corresponding author. E-mail:jianhui.lian@ynu.edu.cn}
}

\usepackage{lineno}

\begin{document}

\maketitle

{\bf Canonical theory predicts galaxies grow “inside-out”, producing their observed negative radial age {gradient}. This picture is challenged by galaxies’ `U-shaped' {colour} profiles---indicating reversed age {gradient}---explained by either outside-in formation or radial migration. Extragalactic observations cannot disentangle these {two possibilities}, but temporally and spatially resolved observations in the Milky Way {offer} a solution.
Here we report a more complex U-shaped age profile of the Milky Way extending to 20 kpc, featuring an outer positive gradient followed by an age plateau of $\sim${5}~billion year beyond {12} kpc. 
Age and chemical abundance distributions of outer disk stars rule out outside-in formation and confirm radial migration as the primary driver {of the outer positive age gradient and plateau}. Our results suggest local star formation {in the Galaxy} truncates {around 12} kpc 
and radial migration has expanded the Milky Way far beyond its native star formation regime out to 20 kpc---a growth mode likely common to disk galaxies. The Milky Way thus provides a critical template to understand disk assembly in external galaxies, improving our understanding of galaxy growth.}

Galaxies usually exhibit {a} negative { radial age} gradient, meaning the inner regions formed earlier than the outskirts, {consistent with inside-out growth}  \citep{sanchez2014-califa,goddard2017b}.  Deep imaging {and spatially resolved spectroscopic} data reveal, however, a reversed positive {colour and age} {profiles} in the very outer {parts of} local  \citep{deJong2007,azzollini2008,bakos2008,williams2009,yoachim2010,zheng2015,ruiz-lara2017} and distant disk galaxies \citep{yu2025}, implying a different assembly history of galactic outer disk compared to the inner part. So far, it {has been} difficult to {obtain a complete} picture of this complex assembly history based on {extragalactic} observations alone, {as these observations} still face challenges to resolve the underlying stellar populations of different ages and chemical {abundances}.    

{In contrast, the} Milky Way serves {as} an ideal laboratory for studying the formation and evolution of a galaxy, thanks to our proximity, {which enables fully resolved} observations of individual stars. Recent massive stellar spectroscopic surveys have delivered measurements of chemical abundances, {ages, {distances,} and radial velocities} for hundreds of thousands of stars beyond the solar vicinity. These spectroscopic surveys combined with Gaia astrometric data {allow for a} direct reconstruction of the evolution history of the {Galaxy as a whole} and each of its components. 
Specifically, the Galactic outer disk has been examined {in terms of} its stellar structure, dynamics, and chemical properties \citep{chen2019,mcmillan2022,tian2024}, yet {a direct connection to the {extragalactic} outer disks remains lacking}.  

Here we present {a} study of the Milky Way's integrated age profile from the bulge to the outer disk to obtain an unbiased view of our Galaxy's assembly history and to bridge the gap between Galactic and extragalactic studies. We use data from the APOGEE and LAMOST stellar spectroscopic surveys, leveraging their wide spatial coverage far beyond the solar vicinity (Methods). In particular, the APOGEE survey, which operates in the near-infrared and {is} thus less affected by dust extinction, has mapped the Milky Way contiguously from the bulge to the outer disk, thus allowing a more comprehensive study of the Galaxy's formation and evolution. Given this unique advantage, we have carefully calculated and corrected for the selection function for the APOGEE survey to reconstruct the intrinsic {three dimensional (3D)} {luminosity and mass} density distribution of mono-age populations \citep{lian2022b} (Methods). After integrating the 3D density distribution into 1D {radial {surface} density profile}, we {use} the {reconstructed surface} density as {weights} to derive the {light- and mass}-weighted average age at each {radius} as presented in Figure~\ref{age-prof}. For reference, we also include the arithmetic average age profiles calculated from the selected raw samples.

The integrated age {profiles are} more accurate, albeit suffering {from} large stochastic uncertainties, because {reconstructing the intrinsic density distribution} requires binning the sample into {high} dimensions {of 3D position and chemical abundances} to reconstruct the intrinsic density distribution. In contrast, the raw average profiles seem more precise because of better {statistics} with coarser binning, but are less accurate {due to} potential {selection effects}. 
{The light-weighted average age is systematically younger than the arithmetic average of the raw sample, with the maximum difference reaching $\sim$1.8~Gyr at $R\sim6-7$~kpc. This difference is mainly due to the survey strategy that leads to less complete observations close to the disc plane where stars are relatively young and bright \citep{Stone-Martinez2025}. It highlights the importance of the correction for the selection function. }

Both the integrated {(i.e., selection-function-corrected)} and {raw} average age profiles of the Milky Way appear rather complex. At Galactocentric radius {$R<10$}~kpc ($\sim$1.7~effective {radii} ${\rm R_{e}}$ \citep{lian2024b}), the Galaxy exhibits a negative age gradient---consistent with typical local disk galaxies  \citep{goddard2017b,lian2018b}, suggesting {that} the main body of the Milky Way and disk galaxies generally grow inside-out \citep{frankel2019}. Beyond this radius, the age gradient flattens, then reverses to {become} positive between {12 and 15}~kpc, before flattening again to a $\sim${5}~billion year plateau between {15} and 20~kpc. This complex profile (outer positive gradient  +  plateau) is observed in APOGEE data both with {and} without selection function correction, {{across} different age catalogs \citep{Stone-Martinez2025}}, and confirmed independently by LAMOST data. 
{Recent studies have also reported older ages in the outer disk \citep{lian2022a} and `U-shaped' age profile \citep{fiteni2026} in the Milky Way, but the selection function was not accounted for in those analyses, which hinders direct comparison with extragalactic observations.} 

While a positive {average} age gradient {has recently been revealed in} the {outer} Milky Way, 
analogous `U-shaped' {colour} {and age} profiles have been observed in the outskirts of local and distant galaxies {for more than a decade}  \citep{deJong2007,azzollini2008,bakos2008,williams2009,yoachim2010,yoachim2012,ruiz-lara2017,yu2025}. This implies a common assembly history of Galactic and extragalactic outer disks. With this reversed age gradient, we predict that the Milky Way also {exhibits} U-shaped {colour} profiles, { which turn out to have a more pronounced U-shape {compared} to nearby galaxies {with down-bending surface brightness profiles and stellar mass $M_*>10^{10}{\rm M_{\odot}}$} (Methods). These reconstructed {colour} profiles {serve as good} {references} to {bridge} the Milky Way with the broader galaxy population and {to} identify Milky Way-like systems. Notably, galaxies with such U-shaped {colour} profiles often exhibit a down-bending broken density {profiles} \citep{pohlen2006}, where the density decreases more rapidly in the outer part. Both breaks occur at similar {radii}, suggesting the same origin of the broken {colour} and density profiles. Interestingly, this is also true in the Milky Way: our Galaxy exhibits a steeper density profile at $R>12$~kpc \citep{lian2024b}, {consistent with} the break radius in the age profile.  

The origin of the `U-shaped' age and {colour} profiles {remains} uncertain, with two leading scenarios: ex-situ origin via radial migration, or in-situ outside-in formation. 
In the radial migration scenario, older stars have more opportunities to interact with galactic substructures (e.g., bar and spiral arms) and migrate farther \citep{sellwood2002}, naturally producing a positive age gradient in the outer disk where local star formation is minimal. This scenario has been suggested to explain the `U-shaped' {colour} {and age} profiles in galaxies \citep{yoachim2012,ruiz-lara2017,yu2025} {and the Milky Way \citep{lian2022a,fiteni2026}}. 
In the outside-in formation scenario, the radial extent of star formation decreases over time---also predicting a positive age gradient. This phenomenon has been suggested to occur in galaxies in the local Universe \citep{bernard2007,perez2013,pan2015,wang2017-outin} and {in} galaxy formation simulations \citep{grand2018}.

While extragalactic observations are not able to differentiate these two possible explanations, the temporally and spatially resolved observations in the Milky Way offer a solution. 
{Specifically, because radial migration does not alter the intrinsic properties of stars, and because stars are born with radial age and metallicity gradients, the presence or absence of radial migration leads to observable differences in the stellar age–metallicity distribution.}
Figure~\ref{schematic} illustrates the distinctive predictions of radial migration and outside-in formation scenarios in the differential age-[Fe/H] distribution of outer disk stars as a function of radii. The radial migration scenario predicts that the outer disk (dominated by migrated stars) retains the age--[Fe/H] relation of their birth radii, 
with more distant regions concentrated toward older ages and lower metallicities. 
The outermost region of the disk is composed {entirely} of the oldest stars formed at the disk edge where star formation truncates.  
In contrast, the outside-in formation scenario predicts a series of age--[Fe/H] relations at different radii with systematic offset in metallicity. This is because enrichment {in the} Galactic and extragalactic {disks is} inhomogeneous, typically more efficient at smaller radii. This results in a negative metallicity gradient of disk stars at birth, which is ubiquitous in the Milky Way \citep{anders2017,Esteban2018,lian2023} and external disk galaxies \citep{belfiore2017,Curti2020,ju2025}. Consequently, more distant stars are expected to have systematically lower metallicities than coeval inner disk stars.    

Resolved observations of the Milky Way, as shown in Figure~\ref{age-feh}, strongly favor the radial migration scenario. At larger radii, the outer disk ($10<R<20$~kpc) stars observed by APOGEE show distributions increasingly concentrated toward older ages and lower metallicities {following} consistent age--[Fe/H] and [Mg/Fe]--[Fe/H] relations. Notably, the lower envelopes of these distributions overlap across radii---consistent with radial migration predictions. Stars between 16 and 20~kpc exhibit nearly identical, narrow distributions in age, [Fe/H], and [Mg/Fe], indicating the outermost disk consists of a single mono-age, mono-abundance stellar population. This single population has intermediate age ($\sim${5}~billion years old) with the lowest metallicity ([Fe/H]$\sim$-0.6~dex) and {the} highest $\alpha$ abundance ([Mg/Fe]$\sim${0.12}~dex) among low-$\alpha$ {populations}. The flat abundance gradient within this radial range is incompatible with {the} outside-in formation scenario but {is} natural under radial migration---only the oldest stars (also with lowest [Fe/H] and highest [Mg/Fe]) born near disk star formation edge around {12}~kpc can migrate this far. 

Detailed vertical structure analysis of mono-abundance populations reveals that the same mono-abundance population in the locally formed disk within $R<12$~kpc {constitutes} a young thick disk substructure \citep{lian2025a}. 
This young thick disk likely {formed} through a recent {event of gas accretion and disturbance} possibly caused by the first {pericenter} passage of {the} Sagittarius dwarf galaxy progenitor. This offers an intriguing possibility: the {long-distance} ({$>5$}~kpc) migration of the {single-age} population discovered beyond {15}~kpc might be related to the perturbation induced by dwarf galaxy merger. 
Additionally, this population also shares the same chemical abundances {as} the stars in the Anti-Galactic Stream and Monoceros Ring \citep{bergemann2018,qiao2024}, reinforcing the disk origin of these substructures. 

To further testify the radial migration scenario, we constructed a disk model incorporating radial migration (Methods). {This model is more realistic and sophisticated than \citep{lian2022a} by simultaneously modelling the star-forming and migrated disk components and allowing {for} wider migration distance using an analytic description from \citep{frankel2018}.} The model initializes a disk with a negative age gradient {within a transition radius ($R_{\rm trans}$), a flat gradient between the transition and a truncation radius ($R_{\rm trunc}$) where a {radially-dependent} maximum age is set, and no star formation beyond the truncation radius. 
The flat gradient between $R_{\rm trans}$ and $R_{\rm trunc}$ is not physically motivated, but {invoked} to match the {flat bottom of the} observed profile between $10<R<13$~kpc, and the age cap is needed to reproduce the age plateau beyond $R>15$~kpc. {The age cap between $R_{\rm trans}$ and $R_{\rm trunc}$ is assumed to decrease linearly with increasing radius, mirroring the growth of the truncation radius observed in distant galaxies \citep{perez2004,trujillo2005,azzollini2008} and simulations \citep{roskar2008b}. The value {of this age cap} at the present truncation radius $R_{\rm trunc}$, {denoted $A_{\rm max}$,} is treated as a free parameter.} 

{Regarding the star formation history, for simplicity, we assume the} age distribution at each {radius} to be {Gaussian} with constant $\sigma$ width of 3~Gyr in light of observed age distribution in the outer disk. 
{We have tested that introducing {a} more sophisticated age distribution of the in-situ disk does not significantly alter the results because the migrated outer disk is always dominated by the stars born close to the truncation radius (Methods). The initial radial density profile of the model is {assumed to be exponential with a scale length of 2.6~kpc \citep{bland2016}.}
Radial migration is implemented as a Gaussian diffusion-like process following \citep{frankel2018}, with older stars redistributed over wider radii, {up to} a maximum migration distance of 9~kpc. 
{The parameters of this} dedicated disk model are chosen after {a broad} exploration of {different possibilities}. A {complete} exploration of this model is presented in Methods. 

{The fiducial model and its parameter uncertainties are determined based on Markov Chain Monte Carlo simulation (Methods). This model} well reproduces the observed complex age profile as illustrated in Figure~\ref{rm-model}. 
Redistribution of stars born within {12}~kpc produces a flatter negative inner gradient, while the outer disk---populated solely by migrated stars---exhibits a positive gradient {(12--15~kpc) that flattens outward (15}--20~kpc)---in remarkable agreement with observations.  
Our {best-fitted} radial migration model indicates that the Galactic disk's star formation truncates at $\sim${12}~kpc {from $\sim$4~Gyr ago to the present}, but radial migration extends the stellar disk out to at least 20~kpc. 
Signatures of disk truncation at similar radius are observed in density profiles of young stellar populations \citep{hunt2024,lian2024b}.  
{However, it is worth noting that our} model does not rule out occasional low-level star formation beyond {12}~kpc, which could form the observed few young Cepheids \citep{chen2019} and open clusters \citep{hunt2024} in this region. 
{While radial migration is critical in populating the very outer disk beyond 12~kpc, {its role transitions to} radial mixing in the inner disk, where {sustained} in-situ star formation becomes important. As a result, the age gradient of the in-situ disk flattens from $\sim-1.2$~Gyr/kpc to $\sim-0.5$ Gyr/kpc. A {similar} moderate flattening of metallicity gradient due to this mixing effect has also been suggested \citep{lian2022a}. This radial mixing effect {thus underscores the importance of} radial migration {as} an important factor in understanding the birth {gradients} of stellar population properties {in} external galaxies.}

Fortunately, {the} radial migration process is well constrained by the integrated age profile. Only {weak to moderate degeneracies exist between the migration strength and the transition and truncation radius and the age cap at the truncation radius}
(Methods). The best-fitted radial migration strength in this work is {3.23$^{+0.26}_{-0.41}$} kpc, characterizing the $\sigma$ of {the} Gaussian redistribution function at a lookback time of 8~Gyr. This estimate is {consistent with} previous {works} based on the same configuration of radial migration {but focusing on the inner disk} (i.e., 3.0-3.6~kpc in \citep{frankel2018,frankel2020} after {conversion} to the same lookback time). Converting the above migration strength to {an} average migration distance over 10~Gyr, our estimate is {2.89}~kpc, which is {moderately higher than} previous estimates of 2.1-2.4~kpc in galaxy simulations under the same definition  \citep{roskar2008b,silva2021,khoperskov2021}. 
{Our estimate of radial migration strength is subject to considerable systematic uncertainties arising from the assumptions of the disk model (Methods).}
{Nevertheless, the observational} constraint of {the average} age profile on radial migration as presented here can be {translated} to {colour} profiles and extended to external galaxies. 

{The physical mechanisms responsible for the radial migration in the very outer disk {remain} uncertain. While the long-distance migration of mono-age, mono-abundance {stellar populations} out to $R>15$~kpc {may be linked} to impulse migration triggered by close interaction with dwarf galaxies, migration of stars to $R=12-15$~kpc {must} be secular {in nature to produce} the positive age gradient. {The central} bar is unlikely to {drive} migration at such {large} distance. Spiral arms, {however,} are promising candidates. For instance, {the} Outer Scutum Centaurus arm {extends into} the disk beyond 12~kpc \citep{reid2019}.}
The age plateau beyond {$15$}~kpc {imposes} {additional} constraints on the star formation history of the Galactic outer disk. It is incompatible with {a sustained} star formation history over cosmic time, but implies an abrupt {onset} of star formation { at least 4}~billion year ago, {possibly} a result of perturbation caused by galaxy merger {or} close interaction with e.g., {the} Sagittarius dwarf galaxy. 

In this work, we report a complex age profile of the Milky Way galaxy that consists of an inner negative gradient at {$R<12$}~kpc, a positive gradient between {$12<R<15$}~kpc, and a {5}~billion year plateau beyond {15}~kpc. This age profile is even more complex than that {implied} by the U-shaped {colour} profile in external galaxies. Resolved distributions of stars in age and chemical abundances rule out the alternative possibility of outside-in formation and strongly favor a radial migration origin of the outer Galactic disk at $R>${12}~kpc, where local star formation is truncated. 
Notably, the {outermost} disk beyond {15}~kpc consists of a single mono-age, mono-abundance population {that likely} migrated from the disk near {12}~kpc, demonstrating that radial migration expands the Galaxy far beyond its star formation regime---a growth mode possibly shared by many other galaxies. 


{\bf Methods}

{\bf Data.}
This work is the fourth in a series investigating the Milky Way's {integrated} properties. 
We use data from the APOGEE \citep{majewski2017} and LAMOST \citep{zhao2012} stellar spectroscopic surveys to derive robust average profiles. 

For the APOGEE sample, we adopt stellar atmosphere parameters and chemical abundances from the APOGEE main catalog, and stellar ages {and} distances from the {AstroNN} value-added catalog \citep{mackereth2019,leung2019} {released in SDSS-IV Data Release 17 (DR17) \url{https://www.sdss4.org/dr17/} \citep{blanton2017,ahumada2020,jonsson2020}, the {final} and complete data release of {the} SDSS-IV survey}.
The following criteria are applied to select the sample:
\begin{itemize}
    \item log$(g)<3.5$ for giant stars,
    \item ${T_{\rm eff}}>3500$~K to avoid cool stars,
    \item target flag ETRATARG==0 for main survey targets,
    \item height {${|Z|}<2$}~kpc and orbital eccentricity $<0.7$ to ensure disk stars,
    \item spectrum signal-to-noise {ratio} $>$ 50.
\end{itemize}
This yields {200,093} stars for the APOGEE sample. For intrinsic density distribution calculations (within $|Z|<$4~kpc), the height {criterion} is lifted, resulting in {209,194} stars. 

{The observed age distribution, vertical age profile, and vertical number distribution in different radial bins of the selected sample are shown in {Supplementary Figure~1.}
A steep vertical age gradient exists in the disk that quickly flattens towards the inner Galaxy. Vertical age gradient and observed vertical number distribution are two key factors affecting the change in average age induced by the correction for the selection function. At the solar radius, the observed vertical number distribution significantly deviates from the underlying population with underrepresented mid-plane stars, while a steep vertical age gradient {is present}. {Consequently}, the observed sample, which preferentially {misses} mid-plane stars, {overestimates} the average age and the correction for {the} selection function makes a {non-trivial} difference.}

{To further assess the robustness of the results, we test stricter selection criteria that {cover} the parameter space more densely populated by the training set for the age determination, which is adopted from {the} APOKASC-2 asteroseismic age catalog \citep{pinsonneault2018}. The U-shaped age profile persists, albeit with {a} much smaller sample size. Additionally, we compile independent age measurements from \citep{Stone-Martinez2025} for our selected stars in APOGEE DR17. The uncorrected age profile of this sample also exhibits {a} U-shape, although the average age is younger in the inner Galaxy within $R<4$~kpc. Notably, this U-shaped age profile remains when using the whole catalog for Data Release 19 of {ongoing} SDSS-V survey.  
To summarize, these tests confirm that the U-shaped age profile {found} in this work is robust against potential uncertainties in age measurements.}

For LAMOST data, we use red clump (RC) stars, with their stellar ages and distances, 
from the value-added catalog provided by \citep{wang2023} (\url{http://www.lamost.org/dr8/v2.0/doc/vac}). A sample of 187,506 RC stars {is} selected by applying the following criteria:
\begin{itemize}
    \item classified as RC stars with flag `status' == 1 (primary RC) or `status' == 2 (secondary RC),
    \item spectrum signal-to-noise {ratio} $>$ 30.
\end{itemize}

{
{Supplementary Figure~2} shows the age comparison for 7,158 common stars in our selected APOGEE and LAMOST samples. LAMOST ages are systematically higher than APOGEE {ages} by a median value of 1.45~Gyr. {While the offset is near the limit of the surveys' precision, this bias remains a distinct systematic effect whose origin is still unknown.} 

{\bf Selection function calculation and correction.}
A detailed description of the procedures to calculate the APOGEE selection function is provided in \citep{lian2022b}; 
here we briefly summarize key steps and refer readers to the previous paper for more details. 

The selection function in this work is defined as the fraction of observed stars in the underlying population. It contains two major components. The first component is the fraction of underlying stars in the target pool. {Because} this component depends on {the 3D positions} and properties of target stars, we first bin the sample by 3D position, [Fe/H] and [Mg/Fe] and conduct all the following steps for each bin independently. For each bin, we generate mock catalogs using {state-of-the-art} PARSEC isochrones \citep{bressan2012}, the Kroupa initial mass function \citep{kroupa2001}, and 3D dust extinction maps \citep{bovy2016,green2019}. Applying APOGEE's target selection criteria (defined in 2MASS ${\rm (J-Ks)_0-H}$ {colour}-magnitude diagram \citep{zasowski2013,zasowski2017,beaton2021,santana2021})  
to the mock catalogs {yields} the fraction of underlying stars in the target pool, {i.e.,} the first component of the selection function. The second component is the fraction of stars in the target pool that are eventually observed and selected in our observed sample. This component is based on the results derived by \citep{imig2023}. The final effective selection function is a product of these two components. 

Dividing {the} observed stellar density by the effective selection function yields the intrinsic 3D density distribution of mono-abundance populations. We then unfold each mono-abundance population at each position in ($R, Z$) {in age dimension} to obtain the density distribution of mono-age populations. Assuming azimuthal symmetry, we fit {the} vertical density distributions at each radius using a single exponential model to derive surface luminosity {and mass} {densities}, which is then used as {weights} to calculate the {light-weighted} and mass-weighted average age. {In addition}, we compute the luminosity densities in a variety of monochromatic bands, including SDSS $u, g, r, i, z$ in optical, and 2MASS $J, H, K_s$ in near-infrared. The predicted $g-r$ and $g-i$ {colour} profiles are presented in {Supplementary Figure~3.}

To illustrate the connection between the Milky Way and external galaxies, we extract the $g-r$ {colour} profiles for a sample of 71 nearby face-on star-forming galaxies within 50~Mpc {that have down-bending surface brightness profiles and} stellar masses $M_*>10^{10}{\rm M_{\odot}}$ using galaxy image data from Dark Energy Spectroscopic Instrument (DESI) Legacy Imaging Survey. The right panel of 
{Supplementary Figure~3} shows these {colour} profiles in comparison with our prediction for the Milky Way {scaled by their break radius. Down-bending disk galaxies exhibit a more pronounced `U-shaped' colour profile, with {the} average $g-r$ colour at the small and large radius ends remarkably consistent with the Milky Way. However, at the break radius, these galaxies show a $g-r$ colour redder than the Milky Way, possibly due to the extraordinarily steep negative metallicity gradient observed in the Milky Way \citep{lian2023}.} 

{\bf Uncertainties of the integrated average age measurements.} 
{We estimate the stochastic uncertainty of our average age measurements using Monte Carlo (MC) simulation, considering Poisson errors of observed number density in each spatial and abundance bin, the {uncertainties in the} age and distance of individual stars. 
In the MC simulation, we resample the distance of each star based on its uncertainty and rebin the sample. With 100 resamplings, the standard deviation of the observed number density is {taken} as the uncertainty caused by distance uncertainty. Given limited {statistics} in the current sample, Poisson error is generally the {dominant} source of uncertainty with typical values twice {those} introduced by distance uncertainty. A similar Monte-Carlo simulation is{employed} to incorporate the age uncertainty when unfolding the recovered density distribution of mono-abundance populations in the age dimension. {Specifically, we perform} the unfolding step 100 times and use the standard deviation of the density of mono-age populations to represent {the} uncertainty inherited from {the} age uncertainty of individual stars. This uncertainty on the integrated age measurements is also minor {compared} to that propagated from Poisson error.} 

{In contrast to stochastic uncertainty, systematic uncertainties {in the} integrated age measurement are {more} difficult to quantify. They} mainly {arise} from the specific choice of the stellar evolution models and 3D extinction maps. Our previous experiments, {using} different isochrones and extinction maps, suggest $\sim$5\% differences in the measurements of integrated properties (e.g., light-weighted average metallicity \citep{lian2023}). In light of these experiments, we expect that including the systematic uncertainty would mostly affect the uncertainty of the integrated age measurements within {12}~kpc. Beyond {that radius}, stochastic uncertainties are much {larger} due to the lower statistics and {are} likely the dominated {source}.  

{\bf Disk toy model.}
We implement radial migration by redistributing stars assuming a Gaussian diffusion-like process following \citep{frankel2018} :
\begin{equation}
P(R|R_0, t) = C \exp \left( -\frac{(R - R_0)^2}{2\sigma_{\rm rm}^2 t/ t_0 } \right)
\label{eqn:migration}
\end{equation}
where $R_0$ and $R$ {represent} the birth and migration target radius, respectively, $t$ indicates the {lookback} time, {$C$ is the normalization factor to ensure the integrated probability equals one,} and $\sigma_{\rm rm}$, {in {units} of kpc}, denotes {the width of Gaussian redistribution function at a certain lookback time $t_0$, which {characterizes}} the strength of radial migration. {Following \citep{frankel2018}, we set $t_0 = 8$~Gyr.} 
{Supplementary Figure~4} illustrates an example of the redistribution for stars born at 8~kpc.

The {disk} model includes six free parameters: 
\begin{itemize}
    \item initial age gradient ($\nabla_{\rm age,ini}$) within the transition radius ($R_{\rm trans}$),
    \item initial age profile intercept ($d_{\rm int}$),
    \item transition radius ($R_{\rm trans}$) beyond which the initial age gradient is set {to} zero and a maximum age {younger than the age of the Universe} is applied,
    \item truncation radius ($R_{\rm trunc}$) beyond which no {ongoing} star formation is assumed,
    \item maximum age ($A_{\rm max}$) of stars born {at $R_{\rm trunc}$},
    \item radial migration strength ($\sigma_{\rm rm}$).
\end{itemize}
The first {five} parameters define the initial disk and the {last one} {describes} the radial migration process. {The maximum age is assumed to decrease linearly from the age of the Universe at $R_{\rm trans}$ to $A_{\rm max}$ at $R_{\rm trunc}$.} Explorations of these parameters are shown in 
Figure~\ref{rm-model} 
in the main paper {for radial migration strength} and 
{Supplementary Figure~5} here {for the rest parameters}. For reference, we also explored the model with single initial negative age gradient. {Since the initial age gradient of the fiducial model would yield negative age at $R>9$~kpc, a flatter initial age gradient is adopted for the single-gradient and comparison model.}

These variant models show that $\nabla_{\rm age,ini}$ and $\sigma_{\rm rm}$ jointly determine 
the observed inner negative and outer positive age gradients; {$R_{\rm trans}$ and $R_{\rm trunc}$ set the {age} and radius of the minimum, respectively;} $A_{\rm max}$, {$R_{\rm trans}$, and $R_{\rm trunc}$ jointly regulate} the outer plateau {and are partially degenerate with $\sigma_{\rm rm}$. Due to these degeneracies, estimates of radial migration strength derived from fitting the average age profile are subject to systematic uncertainties in the configurations of the maximum age within the disk model. We find that, when adopting a constant age cap between $R_{\rm trans}$ and $R_{\rm trunc}$, a higher $A_{\rm max}$ is required at $R_{\rm trunc}$. This provides more relatively old stars that can migrate beyond $15$~kpc, thereby lowering $\sigma_{\rm rm}$ to $\sim2.5$, a value consistent within $1\sigma$ uncertainty . }

To {determine} the best-fitted model and uncertainties of the parameters, we apply Markov Chain Monte Carlo analysis to the observed and predicted age profiles using {the} \texttt{emcee} Python package \citep{foreman2013-emcee}. To better replicate the flat bottom and positive age gradient of the observed age profile, we manually increase the weight of the profile at $R>10$~kpc. The posterior distribution of model parameters is presented in {Supplementary Figure~6.}
The fiducial model's parameters are assumed to be the {median} (i.e., 50$^{th}$ percentile) of their posterior distribution. The lower and upper uncertainties are then assumed to be the {differences between the 16$^{th}$ and 84$^{th}$ percentiles and the median}. {According to the best-fitted model, the truncation radius of the Milky Way expanded from 7.6~kpc at 13.7~Gyr ago to 12~kpc at 3.9~Gyr ago, broadly consistent with {estimates of the} break radius expansion of disk galaxies over cosmic time \citep{azzollini2008}.} 

{{To test the assumption regarding the} star formation history, we have explored models {with a} constant $\sigma$ width of 6~Gyr of the initial age distribution or {radially} dependent width that decreases from 6~Gyr at $R=0$~kpc to 3~Gyr at $R_{\rm trunc}$.  The posterior {distributions of the parameters in} these test model are generally consistent with the fiducial model within $1\sigma$ uncertainty.   
This consistency suggests that our results are robust {with respect to the assumed} star formation history. }

{{Aside from} the description of radial migration, the model constructed here {differs from that of} \citep{frankel2018} {in order} to focus on the observed average age profile. On {the} one hand, we do not include metallicity {so as} to avoid {the} complexity {of} chemical enrichment history. On the other hand, we {extend} the simulation to the very outer disk where the positive age gradient is observed. This feature provides the critical constraint on radial migration in this work, {whereas} in \citep{frankel2018} {radial migration} is constrained by the radially dependent age--metallicity distribution. }

{\bf Data Availability.} {All data presented in this work are available in the public repository \url{https://github.com/lianjianhui/Sourece-data-for-MW-age-prof-paper.git}.} 

{\bf Acknowledgement.}
We are grateful to the reviewers for their constructive comments and suggestions which have significantly improved the robustness and clarity of the paper. J.L. thanks Zhi Li, Zhaoyu Li, Zhaozhou Li, and Hao Tian for insightful discussions. 

Funding for the Sloan Digital Sky Survey IV has been provided by the Alfred P. Sloan Foundation, the U.S. Department of Energy Office of Science, and the Participating Institutions. SDSS-IV acknowledges
support and resources from the Center for High-Performance Computing at the University of Utah. The SDSS web site is www.sdss.org.

SDSS-IV is managed by the Astrophysical Research Consortium for the 
Participating Institutions of the SDSS Collaboration including the 
Brazilian Participation Group, the Carnegie Institution for Science, 
Carnegie Mellon University, the Chilean Participation Group, the French Participation Group, Harvard-Smithsonian Center for Astrophysics, 
Instituto de Astrof\'isica de Canarias, The Johns Hopkins University, Kavli Institute for the Physics and Mathematics of the Universe (IPMU) / 
University of Tokyo, the Korean Participation Group, Lawrence Berkeley National Laboratory, 
Leibniz Institut f\"ur Astrophysik Potsdam (AIP),  
Max-Planck-Institut f\"ur Astronomie (MPIA Heidelberg), 
Max-Planck-Institut f\"ur Astrophysik (MPA Garching), 
Max-Planck-Institut f\"ur Extraterrestrische Physik (MPE), 
National Astronomical Observatories of China, New Mexico State University, 
New York University, University of Notre Dame, 
Observat\'ario Nacional / MCTI, The Ohio State University, 
Pennsylvania State University, Shanghai Astronomical Observatory, 
United Kingdom Participation Group,
Universidad Nacional Aut\'onoma de M\'exico, University of Arizona, 
University of Colorado Boulder, University of Oxford, University of Portsmouth, 
University of Utah, University of Virginia, University of Washington, University of Wisconsin, 
Vanderbilt University, and Yale University.

This work makes use of data obtained with the Dark Energy Spectroscopic Instrument (DESI). DESI construction and operations is managed by the Lawrence Berkeley National Laboratory. This research is supported by the U.S. Department of Energy, Office of Science, Office of High-Energy Physics, under Contract No. DE–AC02–05CH11231, and by the National Energy Research Scientific Computing Center, a DOE Office of Science User Facility under the same contract. Additional support for DESI is provided by the U.S. National Science Foundation, Division of Astronomical Sciences under Contract No. AST-0950945 to the NSF’s National Optical-Infrared Astronomy Research Laboratory; the Science and Technology Facilities Council of the United Kingdom; the Gordon and Betty Moore Foundation; the Heising-Simons Foundation; the French Alternative Energies and Atomic Energy Commission (CEA); the National Council of Science and Technology of Mexico (CONACYT); the Ministry of Science and Innovation of Spain, and by the DESI Member Institutions. The DESI collaboration is honored to be permitted to conduct astronomical research on Iolkam Du’ag (Kitt Peak), a mountain with particular significance to the Tohono O’odham Nation.

{\bf Funding Statement.}
J.L. discloses support for the research of this work from National Natural Science Foundation of China (No.12473021), National Key R\&D Program of China (No. 2024YFA1611600), Yunnan Province Science and Technology Department (No. 202105AE160021), Key Laboratory of Survey Science of Yunnan Province (No. 202449CE340002).

{\bf Author contributions. }
J.L. developed the initial idea, conducted the analysis of APOGEE and LAMOST data, and wrote the manuscript. Q.S. constructed the radial migration model and performed the relevant analysis. {Y.Z. contributed to the colour profiles of external galaxies.} 

{\bf Competing Interests Statement.} The authors declare no competing interests. 


\begin{figure*}
    \centering	
    \includegraphics[width=14cm]{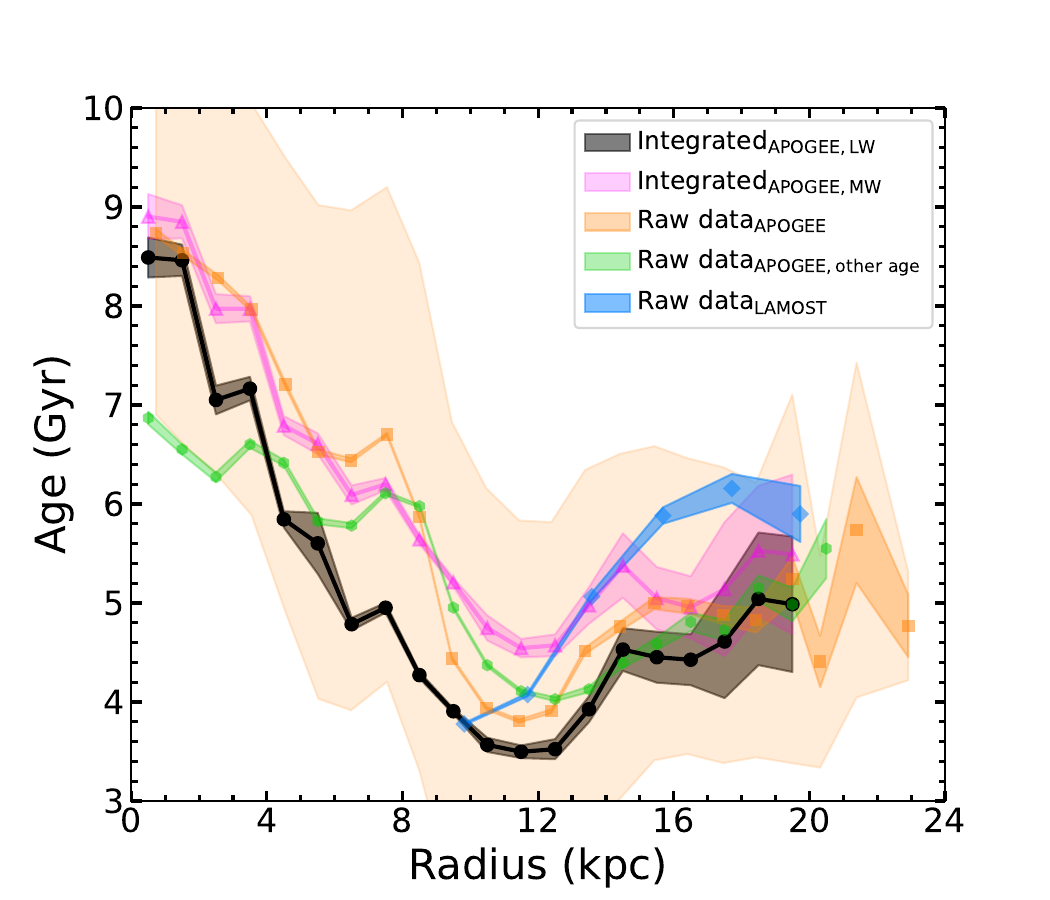}
	\caption{Average age profiles of the Milky Way. 
    Orange squares, {green hexagons, and} blue diamonds show the arithmetic mean age from the APOGEE and LAMOST surveys; black circles {and magenta triangles represent the {light-weighted} and mass-weighted} average age (after selection function correction) derived from APOGEE data. {The colour bands} indicate {the} 1$\sigma$ uncertainties in average age measurements. {The} {1$\sigma$ scatter of age distribution {from the} APOGEE AstroNN data is shown {as the} light orange shaded region}.} 
	\label{age-prof}
\end{figure*} 

\begin{figure*}
    \includegraphics[width=17cm,viewport=60 150 800 440,clip]
    {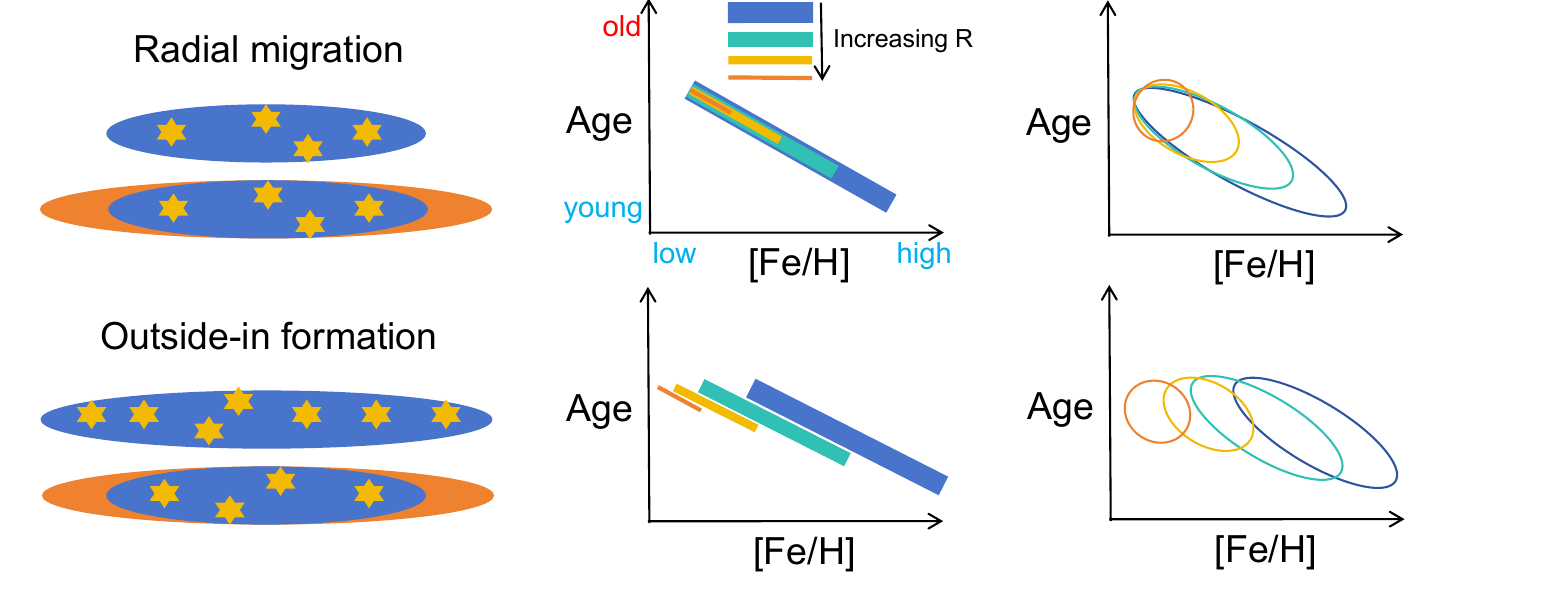}
	\caption{Schematic plot {illustrating} the distinctive predictions of {the} radial migration and outside-in formation scenarios in {the} age--[Fe/H] distribution. In radial migration scenario (top row), the outer disk has minimal local star formation and is dominated by stars migrated from the inner disk (primarily near the star formation edge). Migrated stars retain the age--[Fe/H] {relation} of their birth radius, with more distant regions concentrated toward older ages and lower metallicities. In the outside-in formation scenario (bottom row), outer disk stars form locally, with star formation retreating to smaller radii over time. Due to the negative metallicity gradient {of the Galaxy}, stars formed at larger radii have systematically lower {metallicities} than coeval inner disk stars, producing distinct age--[Fe/H] relations across radii. Line thickness in the middle column denotes surface density as a function of radius.} 
	\label{schematic}
\end{figure*} 

\begin{figure*}
    \includegraphics[width=\textwidth]{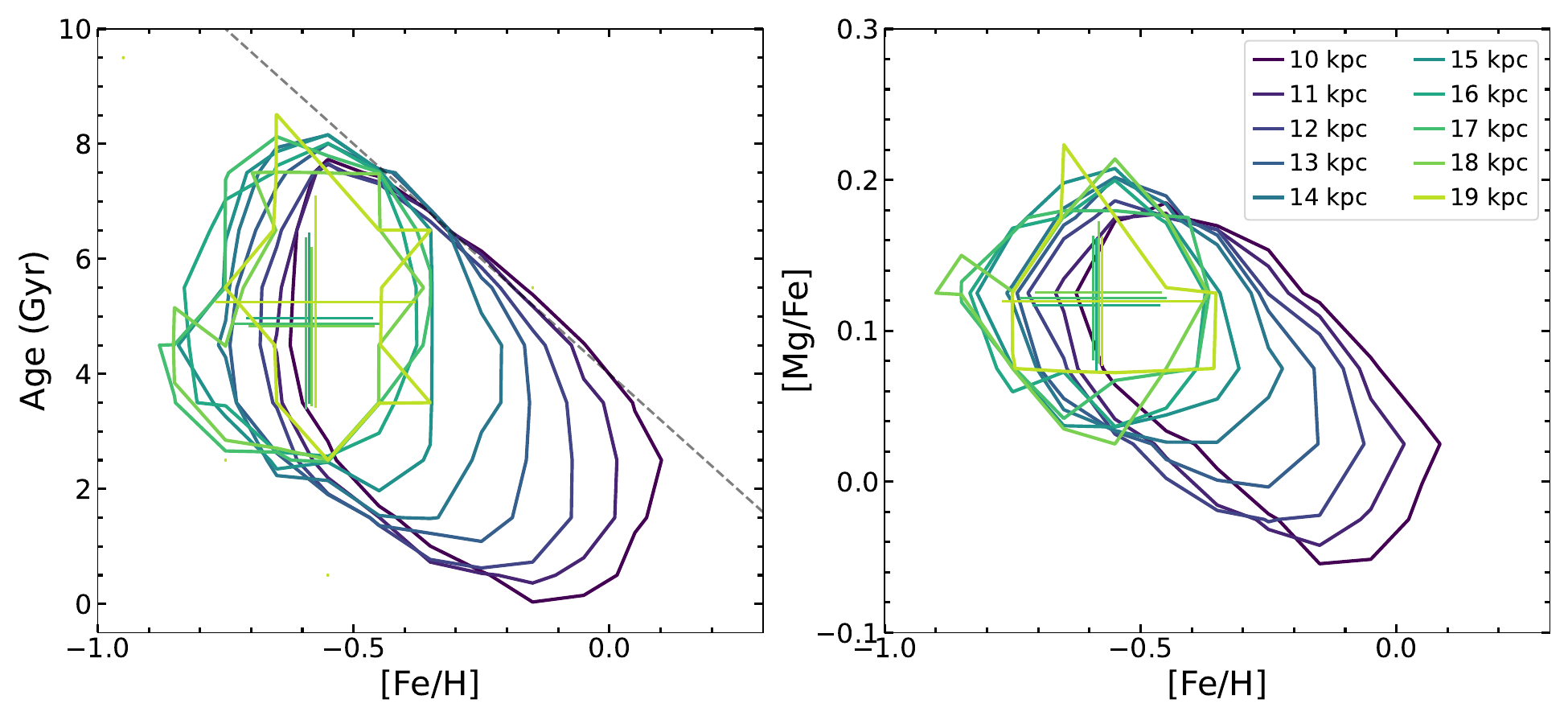}
	\caption{Observed age--[Fe/H] and [Mg/Fe]--[Fe/H] distributions of outer disk stars {from} the APOGEE survey. Contours show the distribution for different radial bins {in} 1~kpc step from 10 to 20~kpc. {The starting radius} of each bin {is} labeled in the legend. 
    The sample is cleaned by removing a small fraction of old stars above the gray dashed line, which belong to a separate older age--[Fe/H] sequence and {likely} migrated from the inner Galaxy \citep{zhang2025}.
    Crosses mark the arithmetic mean age, [Fe/H], and [Mg/Fe] of stars in the outermost bins (16--20~kpc).
    Error bars indicate 1$\sigma$ scatter of the distributions. } 
	\label{age-feh}
\end{figure*} 

\begin{figure*}
    \centering	
    \includegraphics[width=12cm]{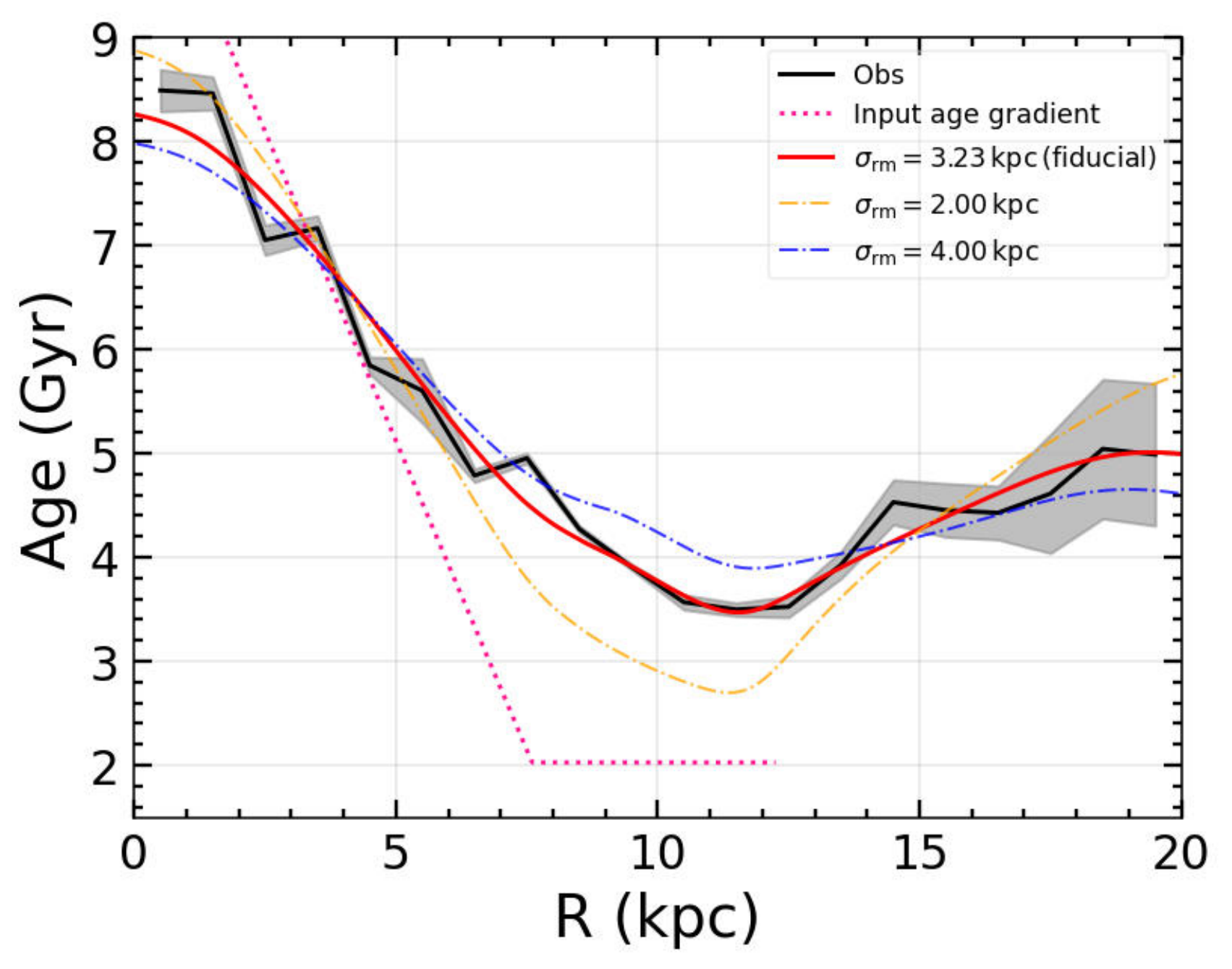}
	\caption{Predicted average age profiles of the radial migration model. The red solid line shows the fiducial model with {{radial} migration strength of {3.23}~kpc.}
    {Blue and orange} dashed and dash-dotted lines show variant models with adjusted {radial migration {strengths}}. A full exploration of the {model's} free parameters is included in Methods. Dotted {magenta} line represents the input piecewise age gradient. The black solid line {and shaded region} {reproduce} the light-weighted age profile {and its $1\sigma$ uncertainty} from Figure~\ref{age-prof} for comparison.} 
	\label{rm-model}
\end{figure*} 

\renewcommand{\figurename}{Supplementary Figure}
\setcounter{figure}{0}

\begin{figure*}
	\centering
	\includegraphics[width=\textwidth]{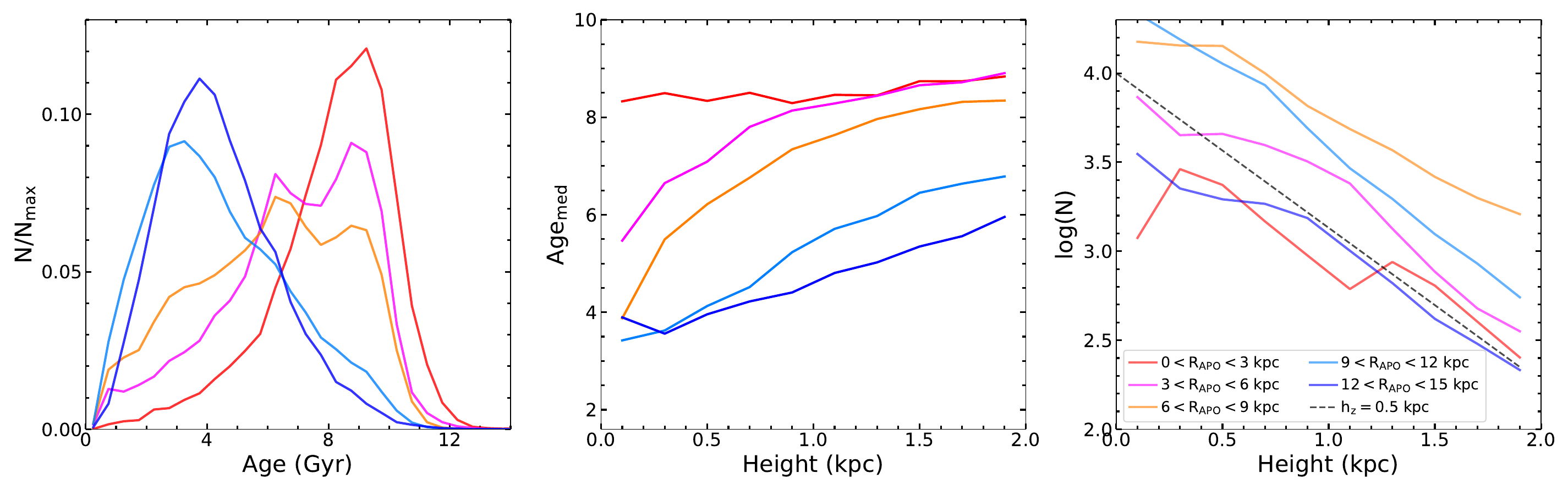}
	\caption{{Age and vertical number {distributions} of {the} APOGEE selected sample. {\sl Left:} Age distribution as a function of radius for APOGEE data using AstroNN ages. {\sl Middle:} Vertical mean age {profiles} at different radii. {\sl Right:} Observed vertical number distribution in different radial bins. {Each colourful line represents the distribution at a certain radial range as indicated in the bottom-right legend in the right panel.} Black dashed line indicates the number distribution (arbitrary normalized value) of {the} underlying population at {the} solar radius reconstructed by \citep{lian2022b} as shown in their Figure~15.} }
    \label{astronn}
\end{figure*}

\begin{figure*}
	\centering
	\includegraphics[width=10cm]{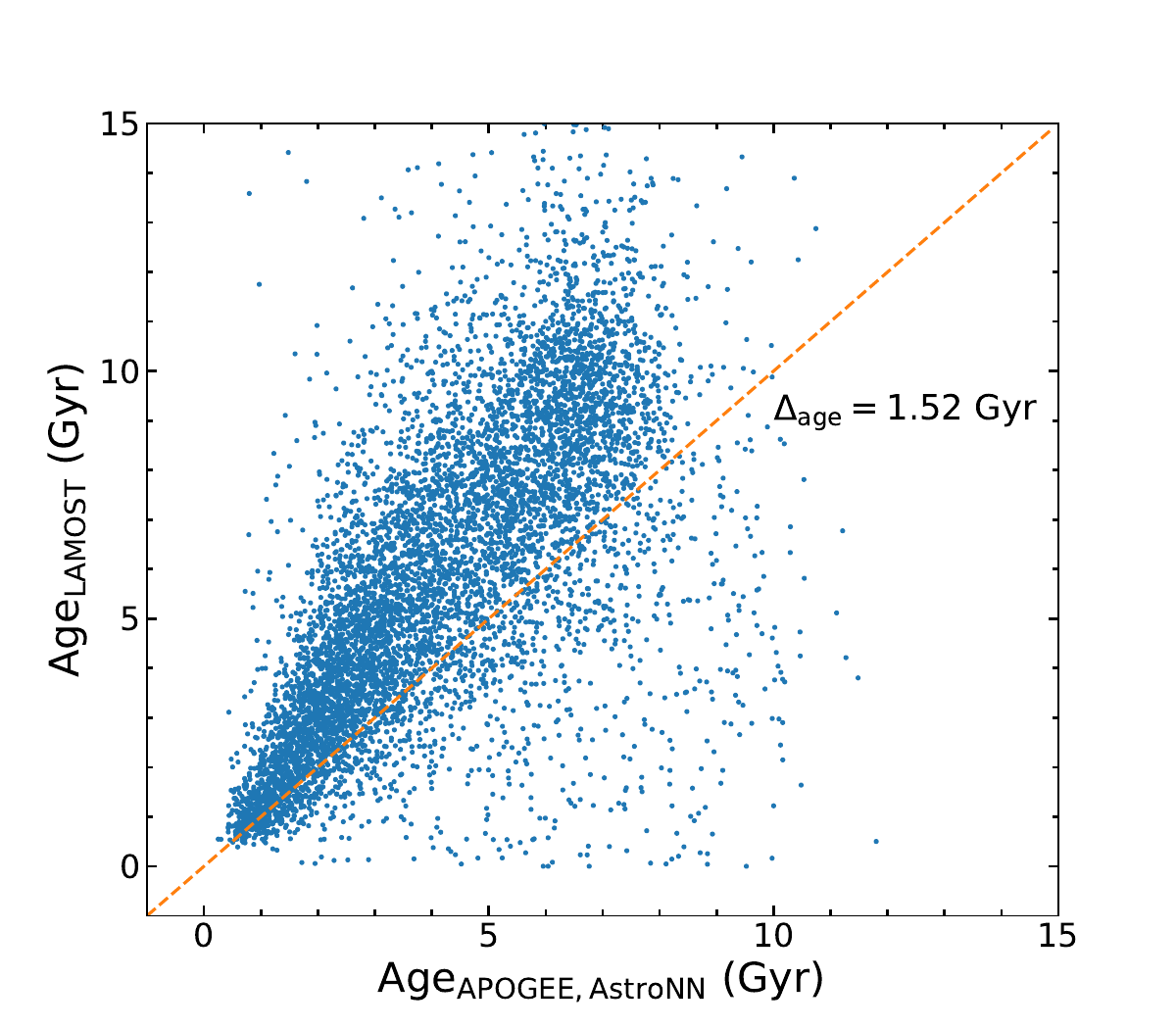}
	\caption{{Comparison of stellar ages between APOGEE/AstroNN and LAMOST for stars in common. The orange dashed line indicates the one-to-one relation. The median value of the difference ($\Delta_{\rm age}$) is indicated in the plot.}}
    \label{age-comp}
\end{figure*}

\begin{figure*}
    \includegraphics[width=\textwidth]{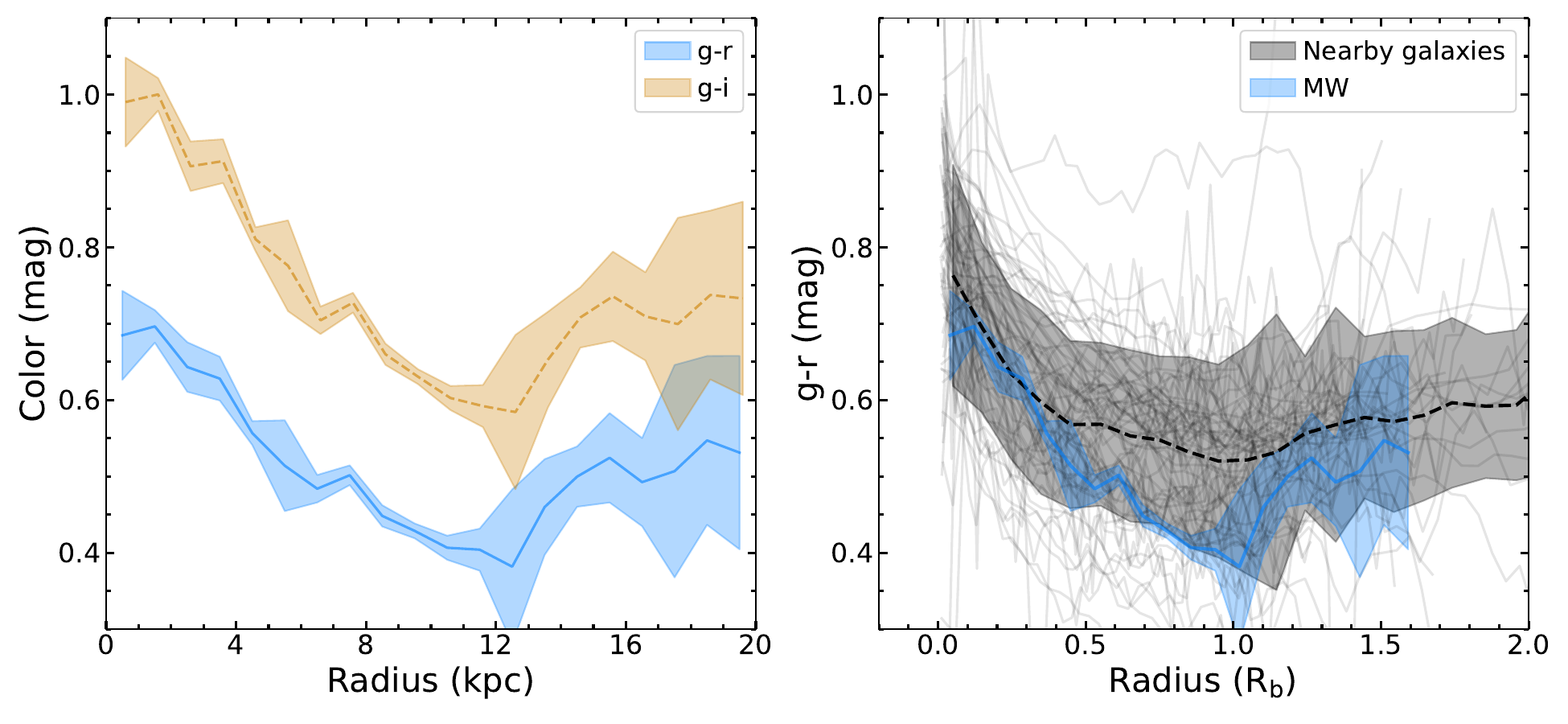}
	\caption{Predicted color profiles of the Milky Way {and comparison with nearby galaxies}. {{\sl Left}: $g-r$ {(blue)} and $g-i$ {(yellow)} colour profiles of the Milky Way predicted based on the reconstructed density distribution of mono-age, mono-abundance populations.} {Colour bands} indicate 1$\sigma$ uncertainties in colour predictions. {{\sl Right:} $g-r$ colour profiles {for 71 nearby face-on star-forming galaxies with down-bending surface brightness profiles and stellar masses $M_*>10^{10}{\rm M_{\odot}}$, normalized to their break radii}. Grey lines indicate the profiles of individual galaxies. Black dashed line and {shaded} region illustrate the median and $1\sigma$ {scatter} of these colour profiles. The Milky Way's $g-r$ profile is the same as {that shown in} the left panel {(blue line)}.}} 
	\label{color-prof}
\end{figure*} 

\begin{figure*}
    \centering	
    \includegraphics[width=10cm]{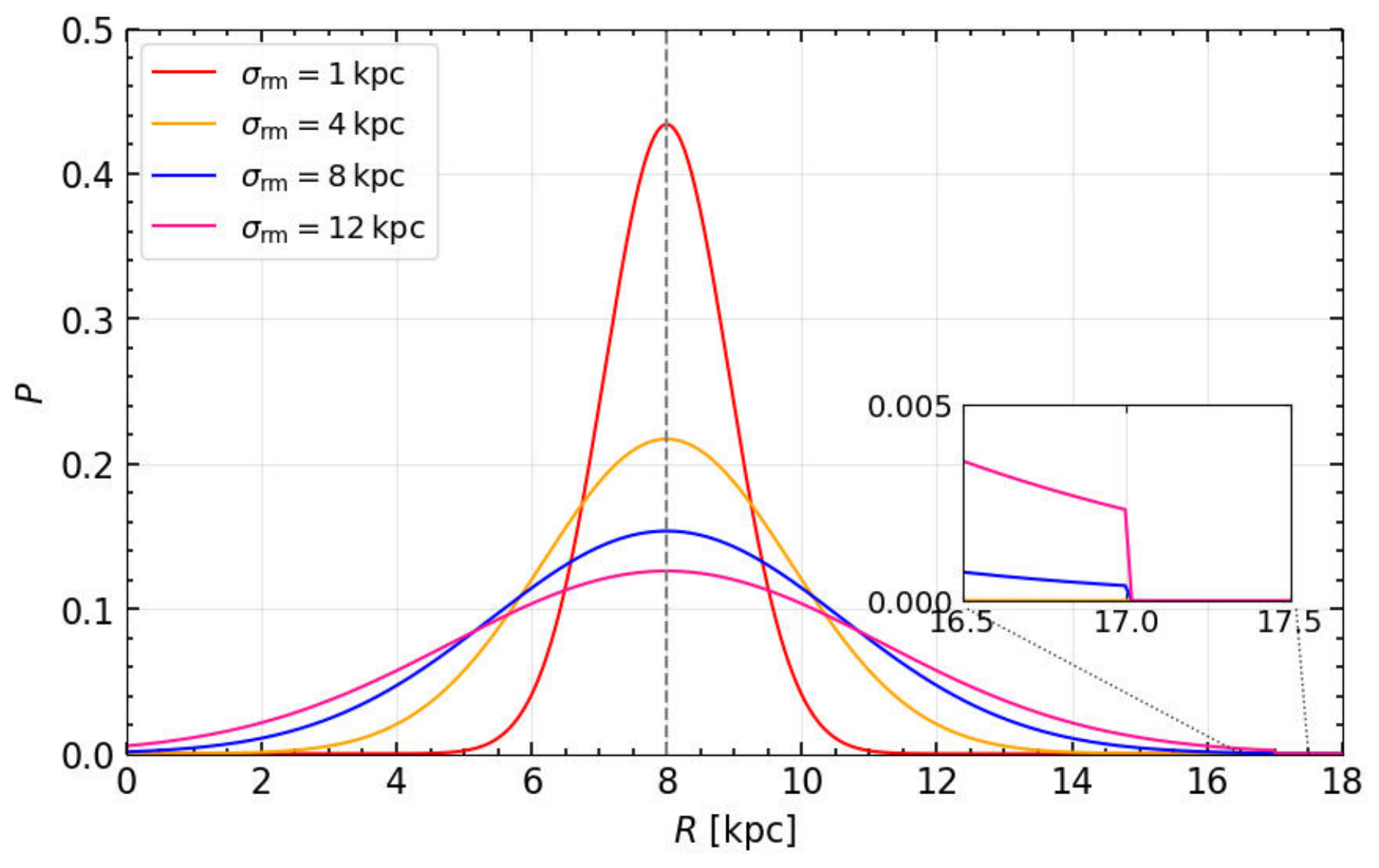}
	\caption{Illustration of radial migration for stars born at 8~kpc, modeled as a Gaussian diffusion process. The x-axis {shows the} radial distance from the Galactic center and {the} y-axis {shows} the probability of redistribution to a given radius. {Each coloured line shows the radial distribution of stars after 8~Gyr for different migration strengths ($\sigma_{\rm rm}$).} The zoomed inset highlights the maximum migration distance (9~kpc) {adopted} in the fiducial model. } 
	\label{gauss}
\end{figure*}

\begin{figure*}
    \includegraphics[width=\textwidth]{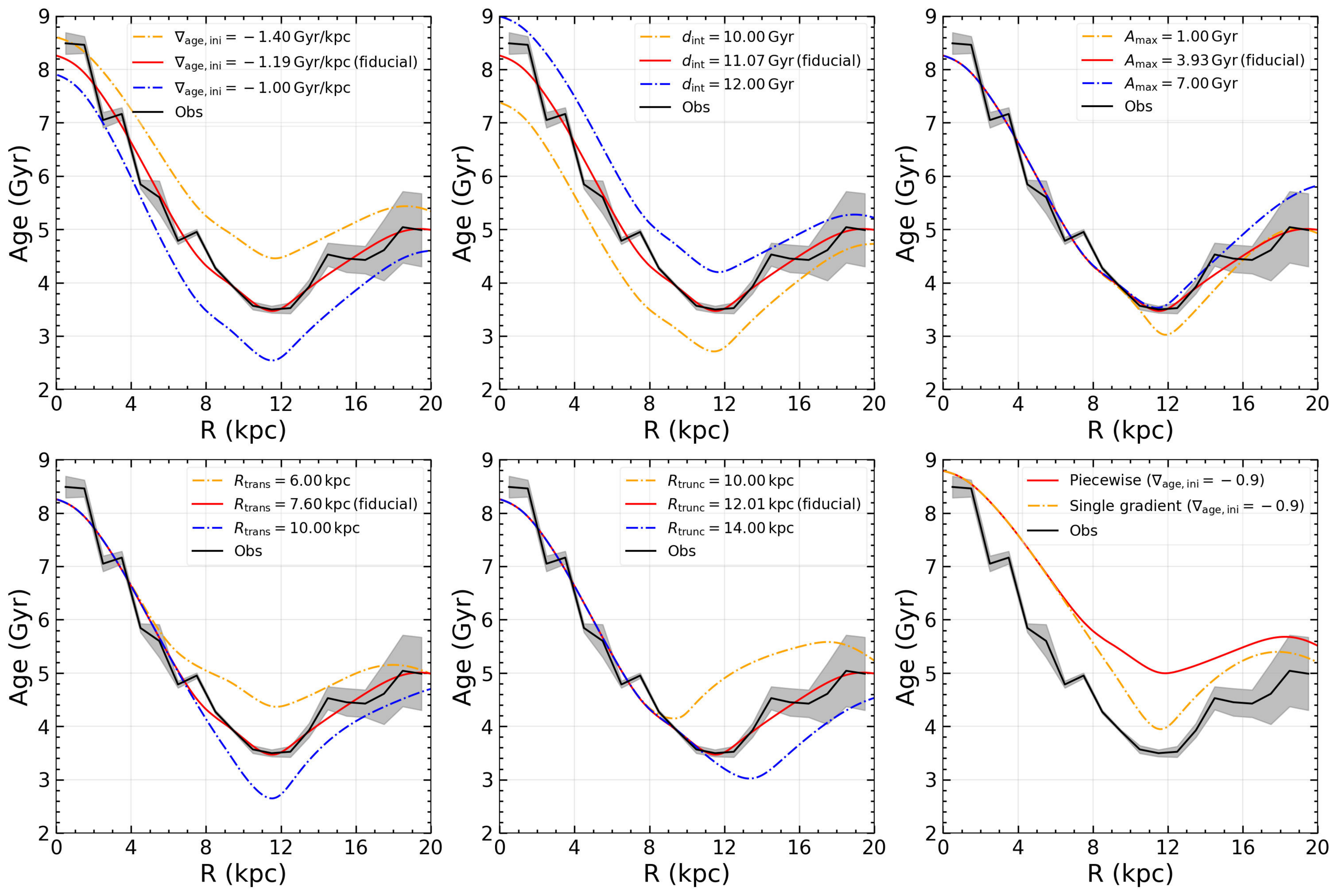}
	\caption{A series of variant radial migration models. Each panel explores the impact of one free parameter on the predicted average age profile. The fiducial model is {shown as} red solid line. From top-left to bottom-right: {(1) initial age gradient ($\nabla_{\rm age,ini}=-1.4, -1.19, -1.0$ Gyr/kpc); (2) initial age profile intercept ($d_{\rm int}=10, 11.07, 12.00$~Gyr); (3) maximum age {at} $R_{\rm trunc}$ ($A_{\rm max}=1, 3.93, 7$~Gyr); (4) transition radius ($R_{\rm trans}=6, 7.6, 10$~kpc);  (5) truncation radius ($R_{\rm trunc}=10, 12.01, 14$~kpc); (6) initial age gradient type (single gradient vs. piecewise gradient). }
    The black solid line and gray shaded region represent the observed luminosity-weighted average age profile and its 1$\sigma$ uncertainty {taken} from Figure 1.}
	\label{rm-free}
\end{figure*}

\begin{figure*}
    \includegraphics[width=\textwidth]{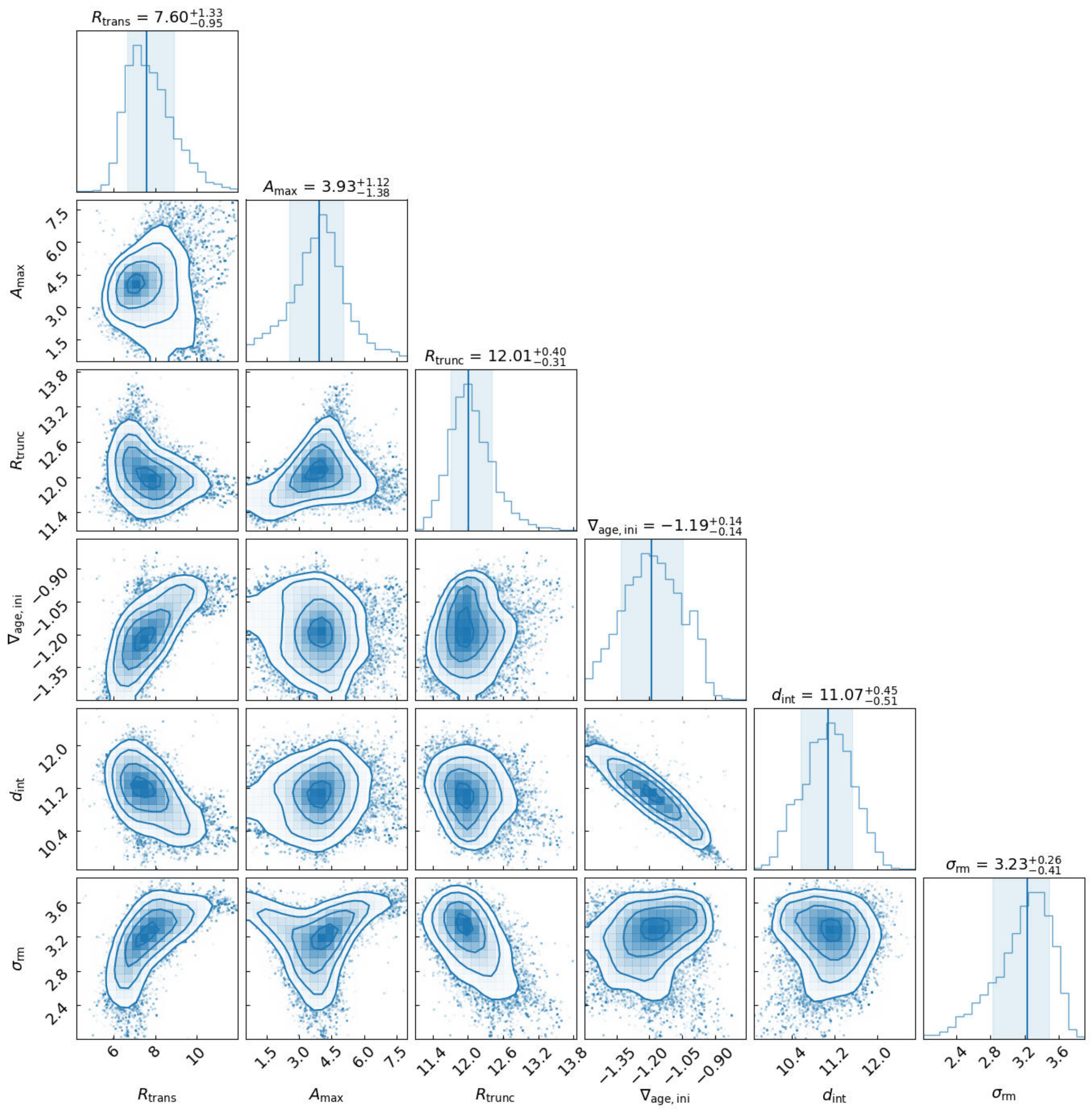}
	\caption{{Posterior {distributions} of the six parameters of the disk radial migration model, {derived from a} Markov Chain Monte Carlo analysis. Blue shaded region in each histogram {represents} the 16$^{th}$ to 84$^{th}$ percentile of the posterior distribution of each parameter. The best-fitted value, {along with the} lower and upper {uncertainties} of each parameter, {is} marked at the top of {each} histogram.}}
	\label{rm-mcmc}
\end{figure*}


\end{document}